\documentclass[twocolumn,aps,showpacs,preprintnumbers,amsmath,amssymb]{revtex4}
\usepackage{graphicx}

\begin{document}
\title{Controlled vortex-sound interactions in atomic Bose-Einstein condensates}
\author{N.G. Parker$^\dag$, N.P. Proukakis$^\dag$, C.F. Barenghi$^\ddag$, and C.S. Adams $^\dag$}
\affiliation{$^\dag$ Department of Physics, University of Durham, South Road, Durham DH1
3LE, United Kingdom}
\affiliation{$^\ddag$ School of Mathematics and Statistics, University of Newcastle, Newcastle upon 
Tyne, NE1 7RU, United Kingdom} 

\begin{abstract}
The low temperature dynamics of a vortex in a trapped quasi-two-dimensional
Bose-Einstein condensate are studied quantitatively. Precession of an off-centred vortex in a dimple trap, embedded in a weaker harmonic trap,
leads to the emission of sound in a dipolar radiation pattern.  Sound emission and reabsorption can be 
controlled by varying the depth of the dimple.
In a shallow dimple, the power emitted is proportional to the vortex acceleration 
squared over the precession frequency, whereas for a deep dimple, periodic
sound reabsorption stabilises the vortex against radiation-induced decay.
\end{abstract}
\pacs{03.75.Lm, 67.40.Vs, 47.32.Cc}
\maketitle

The superfluid nature of weakly-interacting atomic Bose-Einstein condensates (BECs) supports quantized circulation, as 
observed in the form of single vortices \cite{matthews}, 
vortex lattices \cite{abo-shaeer2001}, and vortex rings \cite{anderson}. 
Vortices are fundamental to the understanding of fluid dynamics, signalling the breakdown of ordered flow
and the onset of turbulence. Dilute atomic gases
enable easy control and observation of quantized vortices, complimenting vortex studies
in liquid Helium, superconductors, and non-linear optics. 
Vortices in superfluids are 
subject to both thermal and dynamical instabilities. Thermal dissipation in BECs induces an 
outward motion of the vortex towards lower densities \cite{fedichev}.
Dynamical dissipation is evident in superfluids in the limit of low temperature, as manifested in the temperature-independent 
crystallisation of vortex 
lattices in BEC's \cite{abo-shaeer2002}, and the decay of vortex tangles in liquid Helium \cite{davies}.
In this limit, reconnections and Kelvin 
wave 
excitations of vortex lines leads to dissipation by sound (phonon) emission \cite{vinen,leadbeater}.  Superfluid 
vortices are also unstable to acceleration, 
in analogy to Larmor radiation induced in accelerating charges.  For example, 
corotating pairs 
\cite{vortex_pairs}, and single vortices performing circular motion \cite{lundh,vinen}, within a 
two-dimensional (2D) homogeneous superfluid are predicted to decay via sound emission. 
However, this decay mechanism is not expected to occur in finite-sized BECs due to the sound wavelength being larger than the 
system size \cite{lundh,fetter}. 

In this Letter we show that a vortex in a trapped quasi-2D BEC, precessing due to the inhomogeneous background density, emits 
dipolar sound waves in a spiral wave pattern.  The quasi-2D geometry ensures that the vortex line is effectively rectilinear and that 
Kelvin wave excitations \cite{rosenbusch,garcia_ripoll} are negligible.
This instability is closely analogous to the decay of dark solitons
 in quasi-1D BECs via sound emission due to longitudinal confinement \cite{busch,parker1} .
Quasi-2D `pancake' BECs have recently been created experimentally, using tight confinement in one dimension
\cite{gorlitz,grimm}. Although such systems are prone to strong phase fluctuations, these effects 
are suppressed in the limit of very low temperature \cite{petrov,proukakis}.
In a harmonic trap sound reabsorption occurs, stabilising the
vortex (in the absence of other decay mechanisms), while, in modified trap geometries, sound reabsorption can 
be prevented for times long enough to enable the vortex decay to be observed and probed.  In the latter case, 
the power radiated by the vortex is 
found to be proportional to the vortex acceleration squared and inversely proportional to the precession frequency.  

\begin{figure}[b]
\includegraphics[width=6.8cm]{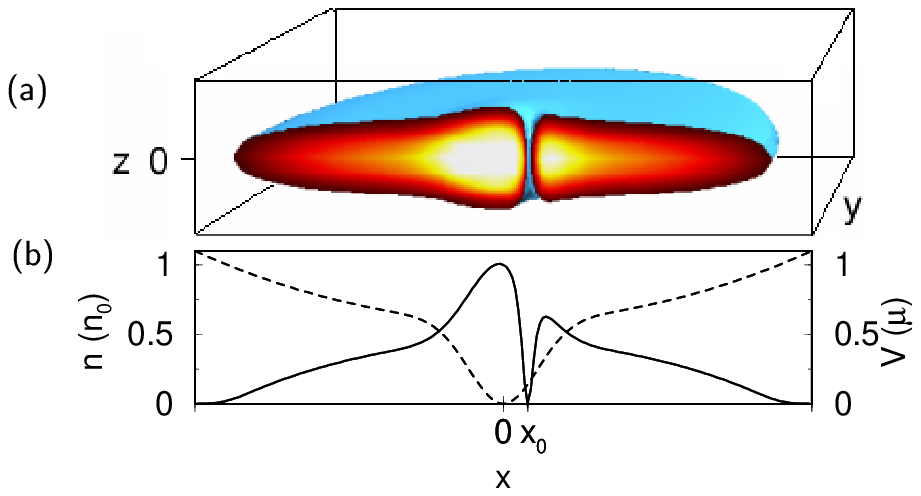}
\caption{(a) Isosurface plot of the atomic density ($n=0.1n_0$, where $n_0$ is the 
peak density) of 
a quasi-2D BEC, confined by the potential of Eq.~(2), 
with a vortex at $(x_0,0)$. In the {\it x-z} plane ({\it y}=0), white 
and black corresponds to high and low density, respectively.
(b) Density (solid 
line) and potential (dashed 
line) along the {\it x}-direction ({\it y}=0,{\it z}=0). }
\end{figure}

Our analysis is based on the Gross-Pitaevskii equation (GPE) describing the mean-field dynamics of a 
weakly-interacting BEC in the limit of low temperature,
\begin{eqnarray}
i\hbar \frac{\partial \psi}{\partial t}=-\frac{\hbar^2}{2m}\nabla^2 \psi+V_{\rm{ext}}\psi+g|\psi|^2\psi-\mu\psi.
\end{eqnarray}
Here $\psi$ is the macroscopic order parameter of the system, $m$ the atomic mass, and $\mu=ng$ is the chemical potential, where 
$n$ is the atomic density.
The atomic scattering amplitude $g=4\pi\hbar^2a/m$, where $a$ is the {\it s}-wave scattering length, is taken to be positive, i.e. repulsive interatomic interactions. The external confining potential $V_{\rm ext}$ is given by,
\begin{equation}
V_{\rm ext}=V_0\left[1-\exp\left(- \frac{2r^2}{\rm{w_0}^2} \right)\right]+
\frac{1}{2}m\omega_r^2 r^2+\frac{1}{2}m \omega_z^2 z^2,
\end{equation}
and consists of a gaussian dimple with waist ${\rm w}_0$ and depth $V_0$ 
embedded within a weaker harmonic trap. This configuration can 
be realised experimentally by focussing a 
far-off-resonant red-detuned laser beam in the centre of a magnetic trap.  
Close to the centre, the gaussian dimple is approximately harmonic with frequency $\omega_{\rm d}=2 \sqrt{V_0}/{\rm 
w_0}$.
For trap parameters, 
$\omega_r=2\pi \times 5$ Hz, $\omega_{\rm d}=20 \omega_r$, 
and $\omega_z= 200 \omega_r$ (we choose $\omega_z \gg \omega_r$ to 
suppress excitation in the 
$z$ direction), the harmonic oscillator time is $\omega_{\rm d}^{-1}=1.6$~ms.
In this case, the timescale of dynamical
instability due to sound emission is much shorter than
the expected thermodynamic vortex lifetime,
which is of the order of seconds \cite{fedichev}.
Assuming a peak density $n_0=10^{14}~{\rm cm^{-3}}$ and a chemical potential $\mu=3.5 \hbar \omega_{\rm d}$, 
a  $^{87} {\rm Rb}$ ($^{23}{\rm 
Na}$) BEC has harmonic oscillator length $l_{\rm
d}=\sqrt{\hbar /(m \omega_{\rm d})}=1.1 (2.1)~\mu$m, and healing length $\xi=\hbar/\sqrt{m \mu}=0.53 l_{\rm d}$.

A singly-quantized vortex, initially at position $(x_0,y_0)$
in the dimple (illustrated in Fig.~1) is expected to
precess around the trap centre, along a path of constant potential, as observed 
experimentally \cite{anderson3}.  This can be 
interpreted in terms of the Magnus force induced by the 
density gradient \cite{jackson,svidzinsky}. 
However, the acceleration of the vortex produces sound emission. 
By varying the depth of the dimple we show how this emission can
be observed and quantified.
Analogous control has previously been demonstrated for dark solitons \cite{parker1}.

The energy of a precessing vortex, for both deep $V_0\gg\mu$ and shallow 
$V_0<\mu$ dimples, is shown in Fig.~2(a).
The vortex energy is monitored by integrating the GP
energy functional,
\begin{eqnarray}
E=-\frac{\hbar^2}{2m}|\nabla \psi|^2+V_{\rm{ext}}\psi^2+\frac{g}{2}|\psi|^4,
\end{eqnarray}
across a `vortex region', defined to be a circle of radius $5\xi$ centred on the core, 
and subtracting off the corresponding contribution of the background fluid.  
Although the vortex energy 
technically extends up to the boundary of the system, 
this region contains $\sim 50\%$ of the total vortex energy at all background 
densities considered here.

For $\omega_z \gg \omega_r$ and providing $l_z \gg a$, where $l_z$ 
is the transverse harmonic oscillator length, the GPE can be reduced to a 2D form with a modified coefficient 
$g_{2D}=g/(\sqrt{2\pi} l_z)$ \cite{petrov,lee}.  
In Fig. 2(a) we compare the full 3D GPE (black lines) with the computationally less demanding 2D GPE
(grey lines), where the 2D and 3D density profiles 
are matched as closely as possible. 
The excellent agreement justifies the use of the 2D GPE
for subsequent results. 

For a deep dimple, the emitted sound waves are confined to the dimple region 
and reinteract with the vortex, and there is no $\it net$ decay of
the vortex energy. The energy oscillations correspond to a beating
between the vortex mode and the collective excitations
of the trapped condensate.  
The beating effect is illustrated in Fig.~2(b), where we plot
the Fourier transform of the vortex {\it x}-coordinate (dotted line) and energy (solid line). 
The two fundamental frequencies, the 
effective trap frequency $\omega_{\rm d}$ and vortex precession
frequency $\omega_{\rm v}$, dominate the position spectrum, while 
the energy spectrum highlights the beat frequencies,
$(\omega_{\rm v}-\omega_{\rm d})$, $(\omega_{\rm v}+\omega_{\rm d})$, 
and higher order combinations.
Similar beating effects are observed for a driven vortex \cite{caradoc_davies},  and 
between a dark soliton and the dipole mode in 
a quasi-1D BEC \cite{parker1}.
In contrast, for a shallow dimple, $V_0<\mu$, the radiated sound escapes,
and the vortex energy decays monotonically (dashed line in Fig.~2(a)).

\begin{figure}[t]
\includegraphics[width=7.8cm]{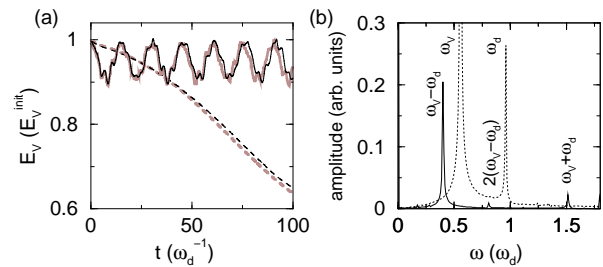}
\caption{(a) 3D (black) and 2D (grey) energy of an off-centered vortex, 
initially located at $(0.53,0)l_{\rm d}$, rescaled by the initial vortex energy. 
$V_0=10\mu$ (solid lines): sound reabsorbed.
$V_0=0.6\mu$, $\omega_r=0$ (dashed lines): sound escapes.
(b) Fourier spectrum of 
the vortex {\it x}-coordinate (dotted line)  
and energy (solid line) for $V_0=10\mu$, using the 2D GPE.}
\end{figure}

In an experiment, the vortex energy can be extracted by
measuring its position. The trajectory of the vortex 
for both deep (solid line) and shallow (dashed) dimples is shown in Fig.~3. 
For $V_0\gg\mu$, the orbit is essentially closed, with the vortex remaining in the effectively harmonic region of the dimple, but features a
small modulation due to the interaction with
the collective excitations of the background fluid.  
In stark contrast, for $V_0<\mu$, the vortex spirals out to lower densities. A similar outward motion has recently been 
predicted for a vortex 
precessing in a harmonic trap modulated by an optical lattice \cite{frantzeskakis}.  
The results presented here are for a homogeneous outer region $\omega_r=0$. Simulations for $\omega_r\neq 0$ are essentially 
indistinguishable up to a time when the sound reflects off the condensate edge and returns to the dimple.  For example, for an 
outer trap
$\omega_r=\omega_{\rm d}/20$, the emitted sound begins to reinteract 
with the vortex at $t \sim 80 \omega_{\rm d}$.  Following this
interaction with the reflected sound, the vortex decay is slowed down, 
but not fully stabilised, due to a dephasing of the sound modes in the outer trap.

Weakly anisotropic 2D geometries yield the same 
qualitative results, with vortex precession now occuring in an 
ellipse, rather than a circle.
In the limit of strong anisotropy, deviations arise as the 
system tends towards the quasi-1D regime,
where vortices are not supported.

The continuous emission of sound waves during the precessional motion is evident by
a close inspection of the surrounding density distribution during 
the course of the decay (Fig.~3, insets). The waves are emitted perpendicularly to the instantaneous 
direction of motion in the form of a dipolar radiation pattern, while the spiralling motion of the 
vortex modifies this into a dramatic swirling radiation distribution, reminiscent of spiral waves often encountered elsewhere in nature 
\cite{spiral_waves}.  
The wavelength of the emitted sound $\lambda \sim 25 l_{\rm d}$ agrees well with the theoretical prediction of $\lambda 
\sim 
2\pi c/\omega_{\rm v}=21.3 l_{\rm d}$ \cite{fetter}, where $c=\sqrt{\mu/m}$ is the speed of sound and $\omega_{\rm v}$ is the vortex precession 
frequency.

\begin{figure}[t]
\includegraphics[width=7.8cm]{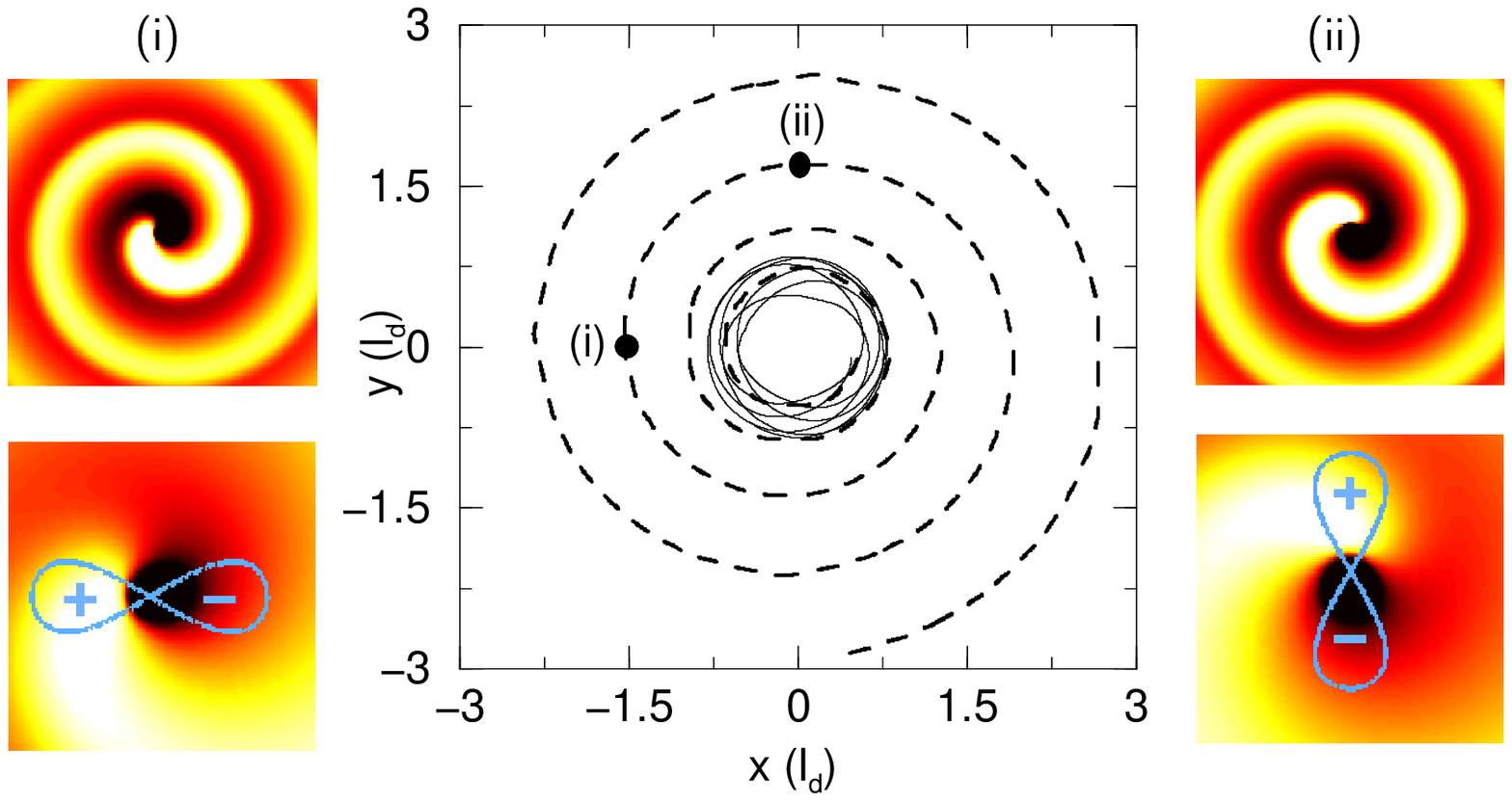}
\caption{ Path of a vortex, initially at $(0.53,0)l_{\rm d}$. 
$V_0 =10\mu$ (solid line): mean radius is constant, but modulated by the sound field. 
 $V_0 =0.6\mu$, $\omega_r=0$ (dashed line): vortex spirals outwards.
Insets: Carpet plots of renormalised density (actual 
minus background density) for $V_0=0.6\mu$ at times $t=$ (i) $61.4$ and (ii) $63.3$ $\omega_d^{-1}$, showing
the emission of positive (white) and negative (black) sound waves of amplitude $\sim 0.01n_0$.
Top: Far-field $[-50,50]\times[-50,50]$. Bottom: Near-field 
$[-14,14]\times[-14,14]$, with schematic illustration of dipolar radiation pattern. }
\end{figure}

The power radiated by the vortex, in the limit of no reinteraction with the emitted sound ($V_0=0.6\mu$, is shown in Fig.~4, as a function of 
time and radius from the trap centre.  Due to contraints on the size of our computational grid, this plot was
mapped out by a few 
simulations, with the vortex being started progressively further from the trap centre.  
This could also be implemented experimentally in order to trace out the vortex 
decay. The curve can be 
understood qualitatively by considering the 
density inhomogeneity that the spiralling vortex experiences: the emitted power increases in line with the local radial density gradient 
up to $r \approx 1.4\xi$, where  the gradient of the 
gaussian potential is a maximum, and subsequently tails off as the trap gradient decreases smoothly to zero.  We have additionally 
considered the case where the dimple is harmonic instead of gaussian, and find the same qualitative results, but with enhanced power 
emission for a particular $\omega_{\rm d}$, due to the larger precession frequency (see Fig.~5, inset).

\begin{figure}
\includegraphics[width=6.8cm]{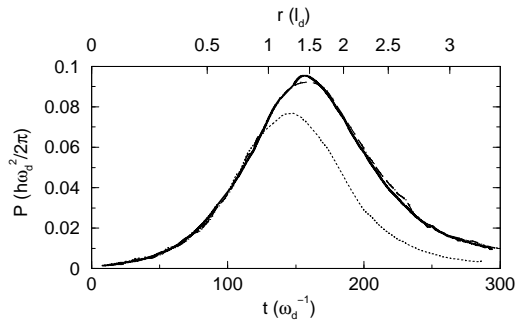}
\caption{Power radiated as a function of radius from the trap centre (top axis) and time (bottom 
axis), as 
calculated from the GP energy functional (solid line). Eq.~(4) with $\beta=6.1$ (dashed line). Acceleration-squared law with constant coefficient 
$26.6 (m/\omega_{\rm d})$ (dotted line).  }
\end{figure}

A 2D homogeneous superfluid can be mapped on to a (2+1)D electrodynamic system, with vortices and phonons 
playing the role of charges and photons respectively \cite{electro}.
By analogy to the Larmor radiation for an accelerating charge and the power emitted from an accelerating dark soliton in a 
quasi-1D BEC \cite{parker1}, we assume the power radiated $P$ by 
the spiralling vortex to be proportional to the square of the local vortex acceleration $a$.   
The coefficient of this relation, $P/a^2$, has been 
mapped out over a range of dimple strengths, as shown in Fig.~5.  Each data point corresponds to the best-fit power coefficient and the average 
vortex precession frequency for that simulation.  In a harmonic trap of frequency $\omega$, the vortex precession frequency is predicted 
to be $\omega_{\rm v}=(3\hbar \omega^2/4\mu){\rm ln}(R/\xi)$, 
where $R=\sqrt{2\mu/m\omega^2}$ is the Thomas-Fermi 
radius of the BEC \cite{svidzinsky}.  For a harmonic trap with a cut-off ($V=V_0$ for $r>r_0$), the 
vortex frequency (Fig.~5 inset, crosses) agrees well 
with 
this prediction.
However, for a 
gaussian dimple of depth $V_0$, $\omega_{\rm v}$ falls
short of this 
prediction, due to the tailing off of the gaussian potential with radius, as shown in Fig.~5 (inset, circles).
Note that there are 
limitations to the range of 
precession frequencies that we can probe, just as would be experienced experimentally: in 
the limit of very tight dimples the vortex escapes almost instantaneously, whereas for very weak dimples, the vortex motion is too
slow for such effects to be systematically studied. The data indicates a strong dependence on the inverse of 
the $\omega_{\rm v}$, suggesting a modified power law of the form
\begin{eqnarray}
P=\beta  m n_0 \xi^2\frac{a^2}{\omega_{\rm v}},
\end{eqnarray}
where $\beta$ is a dimensionless coefficient. 
An equation of this form for
circular vortex motion in a homogeneous 2D fluid
has been obtained by Vinen 
\cite{vinen} using classical acoustics and Lundh {\it et al.} \cite{lundh} 
by mapping the superfluid hydrodynamic equations 
onto Maxwell's electrodynamic equations. 
Both approaches predict a rate of 
sound emission proportional to $\omega_{\rm v}^3 r_{\rm v}^2$, 
where $r_{\rm v}$ is the precession radius, and yield
a coefficient, $\beta=\pi^2/2$.  
Despite the assumptions of perfect circular motion, a point vortex, 
and an infinite homogeneous system, there is remarkable 
agreement with our findings which indicate $\beta \sim 6.3 \pm 0.9$ (one standard deviation), with the variation due to a weak dependence on the 
geometry 
of 
the system.  We believe that the deviation from the predicted 
coefficient arises primarily due to the radial component of 
the vortex motion, which is ignored in the analytical derivations.

\begin{figure}
\includegraphics[width=6.8cm]{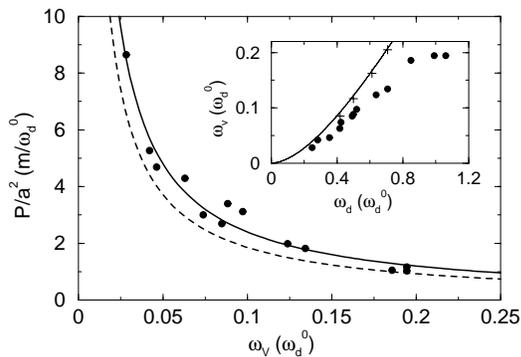}
\caption{Coefficient of an acceleration squared power law, $P/a^2$, for a
vortex, calculated over a variety of trap strengths $\omega_{\rm d}$, as a function of $\omega_{\rm v}$ (circles), along with the analytical 
predictions \cite{lundh,vinen} 
(dashed 
line), and best fit line corresponding to Eq.~(4) with $\beta=6.3\pm 0.86$ (solid line). Here frequency is scaled in terms of $\omega_{\rm d}^0$, 
defined by 
$\mu=3.5\hbar \omega_{\rm d}^0$.
(The gradient of the best fit line in log-log plot is found to be $-1.04$.) 
Inset: Variation of $\omega_{\rm v}$ with trap strength for a gaussian dimple (circles) and harmonic trap with a cut-off (crosses), along with the 
theoretical prediction 
\cite{svidzinsky} (solid 
line, see text).}
\end{figure}

Also plotted in Fig.~4, alongside the power emission from the GP energy functional, are an 
acceleration-squared law (dotted line) and the modifed acceleration squared law of Eq.~(4) (dashed 
line), with the coefficients being chosen to give a best fit.  Both lines give excellent agreement 
until the vortex starts to escape the dimple region at $r \sim 1.4 l_d$. Here the 
vortex frequency, which previously remained roughly constant, starts to decrease due to the 
form of the local density.  This causes the acceleration-squared law to deviate significantly, while 
the $1/\omega_{\rm v}$ term in Eq.~(4) corrects for this deviation, giving excellent agreement throughout 
the decay.

Sound radiation due to acceleration may be
important in the case of turbulent vortex tangles in liquid Helium, 
where evidence suggests that the vortex line 
length $L$ (providing a measure of the energy of the system)
 decays at a rate proportional to $L^2$ \cite{vinen}. In the limit of low temperature, 
this decay is believed to be primarily due to reconnections and
Kelvin wave excitations. We note that, for a system of many vortices, where the acceleration is induced by the surrounding vortex distribution, 
Eq.~(4) would 
also lead to an $L^2$ decay.

In summary, we have shown that a vortex precessing in a trapped 
quasi-2D BEC at low temperature emits dipolar radiation, which becomes 
modified into a spiral wave pattern due to the motion of the 
vortex.  The vortex energy decays at a rate proportional to its 
acceleration squared and inversely proportional to the precession frequency.  
For appropriate trap geometries, the sound emission is 
experimentally observable via the spiralling motion of the 
vortex towards lower densities.  An analogous instability may arise in the case of optical
vortices, which also exhibit a fluid-like motion \cite{rozas}.
For harmonic traps the  vortex 
decay is stabilised by reinteraction with the emitted sound.


\begin{thebibliography}{99}
\bibitem{matthews} M. R. Matthews {\it et al.}, Phys. Rev. Lett. {\bf 83}, 2498 (1999).
\bibitem{abo-shaeer2001} J. R. Abo-Shaeer, C. Raman, J. M. Vogels, and W. Ketterle, Science {\bf 292}, 476 (2001).

\bibitem{anderson} B.P. Anderson {\it et al}., Phys. Rev. Lett. {\bf 86}, 2926 (2001).
\bibitem{fedichev} P. O. Fedichev and G. V. Shlyapnikov, Phys. Rev. A {\bf 60}, R1779 (1999).
\bibitem{abo-shaeer2002} J.R. Abo-Shaeer {\it et al}., Phys. Rev. Lett. {\bf 88}, 070409 (2002).
\bibitem{davies}  S.L. Davis {\it et al}., Physica B {\bf 280}, 43 (2000).
\bibitem{vinen}W.F. Vinen, Phys. Rev. B {\bf 61}, 1410 (2000); W. F. Vinen, Phys. Rev. B {\bf 64}, 134520 (2001).
\bibitem{leadbeater}M. Leadbeater {\it et al.}, 
Phys. Rev. Lett. {\bf 86}, 1410 (2001); M. 
Leadbeater, D. C. Samuels, C. F. Barenghi, and C. S. Adams, Phys. Rev. A {\bf 67}, 015601 (2003).
\bibitem{vortex_pairs} L. M. Pismen, {\it Vortices in nonlinear fields} 
(Clarendon 
Press, Oxford, 1999).
\bibitem{lundh} E. Lundh and P. Ao, Phys. Rev. A {\bf 61}, 063612 (2000).
\bibitem{fetter} A. L. Fetter and A. A. Svidzinsky, J. Phys {\bf 13}, R135 (2001).
\bibitem{rosenbusch}P. Rosenbusch, V. Bretin, and J. Dalibard, Phys. Rev. Lett. {\bf 89}, 200403 (2002).
\bibitem{garcia_ripoll} J. J. Garcia-Ripoll and V. M. Perez-Garcia, Phys. Rev. A {\bf 63}, 041603 (2001).
\bibitem{busch} T. Busch and J. R. Anglin, Phs. Rev. Lett. {\rm 84}, 2298 (1999).
\bibitem{parker1} N. G. Parker, N. P. Proukakis, M. Leadbeater, and C. S. Adams, Phys. Rev. Lett. {\bf 90}, 220401 (2003).
\bibitem{gorlitz} A. G\"{o}rlitz {\it et al}., Phys. Rev. Lett. {\bf 87}, 130402 (2001).
\bibitem{grimm} D. Rychtarik, B. Engeser, H. C. Nagerl, and R. Grimm, preprint, cond-mat/0309536.
\bibitem{petrov} D. S. Petrov, M. Holzmann, and G. V. Shlyapnikov, Phys. Rev. Lett. {\bf 
84}, 2551 (2000).
\bibitem{proukakis} U. Al Khawaja, J. O. Anderson, N. P. Proukakis, and H. T. C. Stoof, 
Phys. Rev. A {66}, 013615 (2002).
\bibitem{anderson3} B. P. Anderson, P. C. Haljan, C. E. Wieman, and E. A. Cornell, Phys. Rev. Lett. {85}, 2857 (2000).
\bibitem{jackson} B. Jackson, J. F. McCann, and C. S. Adams, Phys. Rev. A {\bf 61}, 013604 (2000).
\bibitem{svidzinsky} A. A. Svidzinsky and A. L. Fetter, Phys. Rev. A {\bf 62}, 063617 (2000).
\bibitem{lee} M. D. Lee, S. A. Morgan, M. J. Davis, and 
K. Burnett, Phys. Rev. A {\bf 65}, 043617 (2002).
\bibitem{caradoc_davies} B. M. Caradoc-Davies, R. J. Ballagh, and K. Burnett, Phys. Rev. Lett. {\bf 83}, 895 (1999).
\bibitem{frantzeskakis}P. G. Kevrekidis {\it et al.}, J. Phys. B {\bf 36}, 3467 (2003).
\bibitem{spiral_waves} See I. Biktasheva and V. N. Biktashev, Phys. Rev. E {\bf 67}, 026221 (2003) and references therein.
\bibitem{electro} D. P. Arovas and J. A. Friere, Phys. Rev. B {\bf 55}, 1068 (1997) and references therein.
\bibitem{rozas} D. Rozas, Z. S. Sacks, and G. A. Swartlander Jr., Phys. Rev. Lett. {\bf 79}, 3399 (1997).
\end{thebibliography}
\end{document}